\documentclass[aps,epsfig,preprint]{revtex4-1}
\usepackage{color} 
\usepackage{xcolor}
\usepackage{graphicx} 
\usepackage{epsfig}

\newcommand{\bee}{\begin{equation}}
\newcommand{\ene}{\end{equation}}
\newcommand{\beea}{\begin{eqnarray}}
\newcommand{\enea}{\end{eqnarray}}

\begin{document}
\title{Observation of enhanced absorption of  laser radiation by  nano structured targets in PIC simulations}
 \author{Chandrasekhar Shukla}
  \email{chandrasekhar.shukla@gmail.com}
    \author{Amita Das}
   \email{amita@ipr.res.in}
  \affiliation{Institute for Plasma Research, HBNI, Bhat, Gandhinagar - 382428, India }
\date{\today}
\begin{abstract} 
 It is well known that 
 Brunel's vacuum heating mechanism is operative for laser energy absorption when the target plasma density 
 rises sharply.  For non-relativistic laser intensities and  planar targets  it is also  necessary that the  laser should strike the target at oblique incidence. 
 The  laser electric field at oblique incidence has a  component normal to the surface to 
  extract  electrons from the target in the vacuum region for Brunel's mechanism to be operative. At relativistic laser intensities, oblique incidence is not necessarily required 
   as    the $\vec{J} \times \vec{B}$ force is significant and can extract electrons from  the target even when the laser is at   normal incidence. In this manuscript, the interaction of short and intense laser pulse  with structured  overdense 
 plasma targets have been studied  using  2D particle-in-cell simulations. It is shown that for structured targets  the absorption increases many fold. 
A detailed study and understanding of the absorption process for the structured targets in terms of structure scale length and amplitude have been provided.  
 
  \end{abstract}
 \maketitle 
\section{Introduction} 
The production of high energetic charge particles during the interactions between the intense, short pulse laser with overdense plasma and subsequently, its collimated propagation
inside the target are important
for many application such as fast ignition scheme of inertial confinement fusion (ICF) \cite{tabak_05}, generation of energetic charged particle beams 
\cite{rb,skar,fuchs}, bright source
of X-rays \cite{A,CHEN}, generation of high-order harmonics \cite{U} etc. Therefore, the basic understanding of absorption of laser radiation into overdense 
plasma and manipulation of the same by controlling the laser and  plasma parameters are highly desired. 
In conventional, intense laser-solid interactions, the laser radiation is absorbed by collisionless processes viz vacuum heating \cite{VACUUM}, resonance absorption 
\cite{RESO} and $\vec J\times \vec B$ heating \cite{JCROSSB} etc. However, for short femtosecond (fs) laser pulses and steep density profile of plasma, the laser radiation is 
absorbed by vacuum heating and $\vec J\times \vec B$ mechanism.
In the case of vacuum heating, a p-polarized laser  obliquely incident on a plasma surface has an electric field component normal to plasma surface. 
This normal component of electric field  pulls the electrons 
into the vacuum during one-half of the laser cycle and then returns them back into the target with a quiver velocity in next half laser cycle. 
However, at normal incidence, vacuum 
heating would be absent as there would be no component of the electric field vector normal to the plasma surface. 
The  absorption process will then take place solely  through  $\vec J\times\vec B$ mechanism which is typically small when the laser intensity 
is small so as to have electron quiver  velocity in the electromagnetic field of the laser radiation  non-relativistic.  

Recently, there have been a lot of experimentations with structured targets \cite{m_purvis,g.r.k}.
 There have also  been indications that the laser absorption improves when structured targets are employed \cite{struct1,struct2,struct3,g.r.k}. We explore this question  with the help of 
Particle - In - Cell (PIC) studies in the present manuscript. We treat the structured target as a given profile for plasma density in our simulations. 
The premise for this is based on the fact that the laser intensity at the front will inevitably ionize the target. However, in the case of a short pulse, the ionized 
plasma medium will not have a sufficient time to expand before it encounters the main region of the pulse. Thus the structure of the target will be embedded in the 
density profile of the preformed plasma. 

For our numerical study, we will, therefore,  consider the interaction of the main laser pulse with both planar plasma density targets and those having a 
  specified structure. The simulations clearly show that  for structured targets the laser absorption gets significantly enhanced leading to the 
  heating of the electrons. A detailed study is carried out which illustrates that Brunel mechanism \cite{VACUUM} is operative in a novel fashion for the structured targets 
  wherein the electrons get extracted in the vacuum region from a larger area when the target is structured. 

We have organized this paper as follows. In Section II, we describe the simulation set-up for the PIC simulations.
 In section III simulation results are 
presented which show the enhancement in absorption and the significant increase in the heating of the electrons. 
In section IV the dependence of absorption on structure scale length has been provided. 
Finally, we conclude our 
findings in section V.
\section{Simulations set-up}
\begin{figure}[!htb]
               \includegraphics[width=\textwidth,trim={0 0cm 0cm 0},clip]{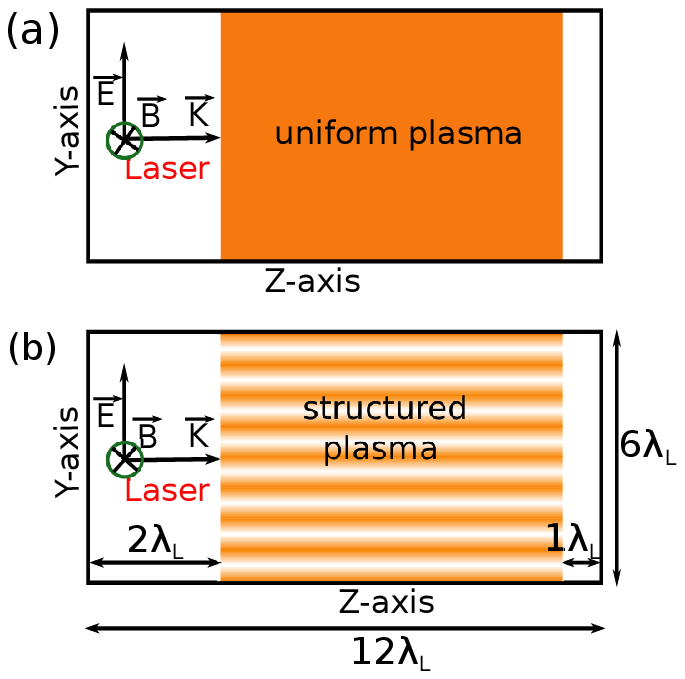} 
               \caption{The schematic of simulation set-up for a p-polarized laser-plasma interactions at normal incidence: electric field of 
             laser is in Y-direction, magnetic field of laser is in "-ve" X-direction, and propagation vector is in Z-direction
             (a) Planar uniform plasma target  (b) Inhomogeneous structured target  }  
                \label{fig1}
        \end{figure}

For our  simulation, a Gaussian laser beam with FWHM of  $3 \mu$m 
 with a top hat temporal
profile of 30 femtoseconds (fs)  enters from the left boundary and interacts with a $2$-$D$ plasma slab. 
The laser is chosen to be 
 p-polarized with its  electric field in the Y-direction and the magnetic field in the "-ve" X-direction as shown in Fig.~\ref{fig1}. The laser propagates in the
Z-direction and interacts with plasma slab at  normal incidence. The size of plasma slab is taken as  $ 6 \lambda $ 
in the transverse direction $\hat{y}$  and $ 9 \lambda$ in the longitudinal direction $\hat{z} $ of the propagation direction of laser light (where $\lambda=1 \mu m$ is laser wavelength). 
There is a vacuum region of $ 2 \lambda$ in front of plasma slab and $ 1 \lambda$ at  the rear of the plasma slab. The absorbing  boundary conditions
have been used for the electromagnetic fields and reflecting boundary conditions have been used for particles.  
  We have used both planar target with the uniform plasma density of $10 n_c$ (where $n_c$ is the critical plasma density for laser) as well a 
  plasma slab with a ripple in its density chosen to be of the form of 
 $ n_{0i}(y) =n_{0e}(y)
  =n_{0}[1+\varepsilon cos(k_{s}y)] $. Here  $l_s = 2\pi/k_s$ represents  density scale length associated with the density inhomogeneity having an 
 amplitude of  $\varepsilon$. The schematic of simulation set-up for both planar as well as inhomogeneous structured target has been shown in Fig.~\ref{fig1}(a) and in Fig.~\ref{fig1}(b) respectively. 
 The mesh size is chosen to be $\delta z$=$\delta y$=0.02 c/$\omega_p$. This resolves the scale length of density inhomogeneity 
 appropriately. Furthermore, $\omega_{p}=\sqrt{4\pi n_{0}e^2/m_{e}}$ is plasma frequency 
 and c/$\omega_p=5\times 10^{-6}cm$ is electron skin depth. The ions, having charge and mass of the proton, are kept at  rest during simulation. 
 \section{Laser absorption}
\begin{figure}[!htb]
               \includegraphics[width=\textwidth]{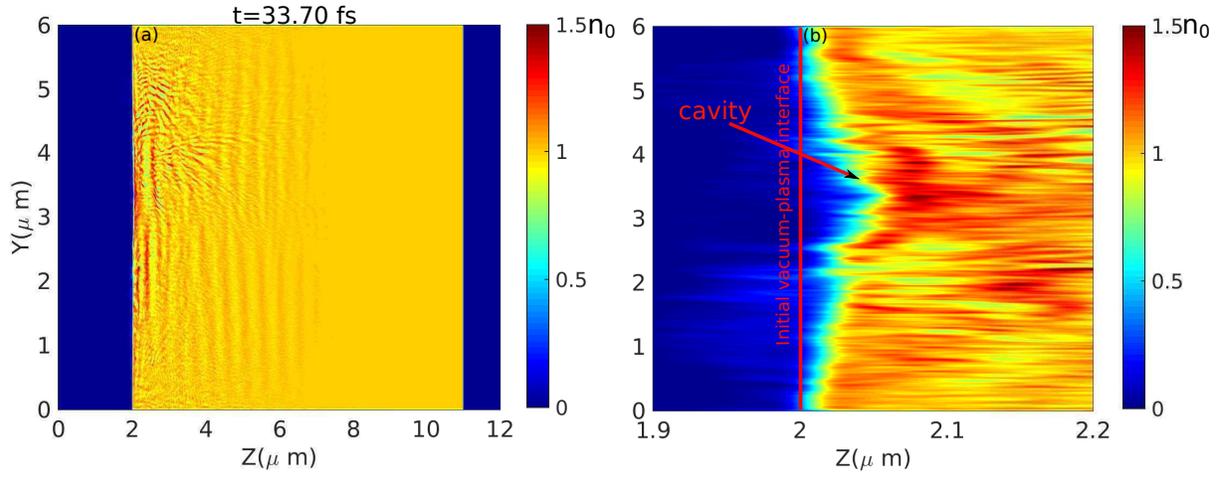} 
              \caption{(a)The electron density [ in unit of $n_0$ ] for case of planar target at
time $t = 33.70$ fs for intensity I$=1\times10^{19}W/cm^{2}$: (b)Zoomed image of (a) where we can see the formation of the cavity due
to evacuation of electrons from high-intensity region to low-intensity region (highlighted
by arrow) }  
                \label{fig2}
        \end{figure}

\begin{figure}[!htb]
               \includegraphics[width=\textwidth]{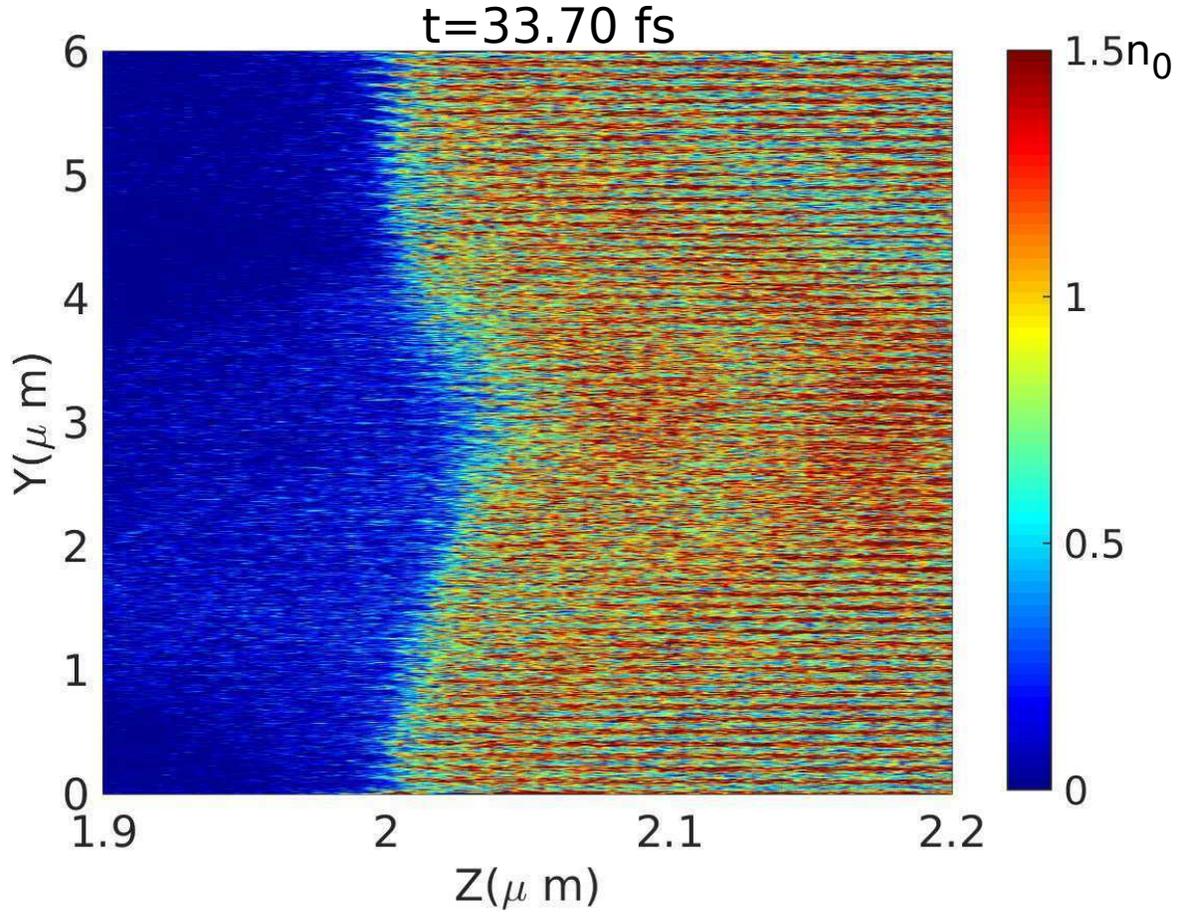} 
               \caption{The electron density [ in unit of $n_0$ ] for structured inhomogeneous target ($\varepsilon=0.5$, $k_{s}=\pi$) at
time $t = 33.70$ fs for intensity I$=1\times10^{19}W/cm^{2}$: the formation of cavity similar to planar homogeneous target can be seen.}  
                \label{fig3}
        \end{figure}

We  consider the case of a low  intensity ( $10^{13} W/cm^2$) laser incident  on both planar homogeneous and inhomogeneous plasma target. The plasma density rise for 
both these cases is steep (compared to the laser wavelength). For the choice of a Gaussian laser profile, the laser intensity varies along the transverse direction.  
 The ponderomotive force felt by the electrons are thus different at the different transverse locations of the target surface. In the central region due to higher 
 laser intensity, the electrons should get  pushed out  towards 
  lower intensity. The depletion of electrons from high-intensity region should form a  cavity. The formation of such  cavity  can be clearly seen 
  for  relativistic  intensities of I$=10^{19} W/cm^2$
 that we have considered in our studies here in Fig.~\ref{fig2}(b) ( which is a zoomed image of Fig.~\ref{fig2}(a)) at time $t=33.70$
  fs) for the homogeneous target. The same structure gets formed for the inhomogeneous target which has been observed Fig.~\ref{fig3}.
  Thus a transverse dimple in the plasma electron density profile  automatically also gets created. 
 However, such  a "caving in" of electron density is very small compared to the laser wavelength. 
 The effect of this small structure formation is negligible on absorption properties which are the issue of main concern here. 

\begin{figure}[!htb]
               \includegraphics[width=\textwidth]{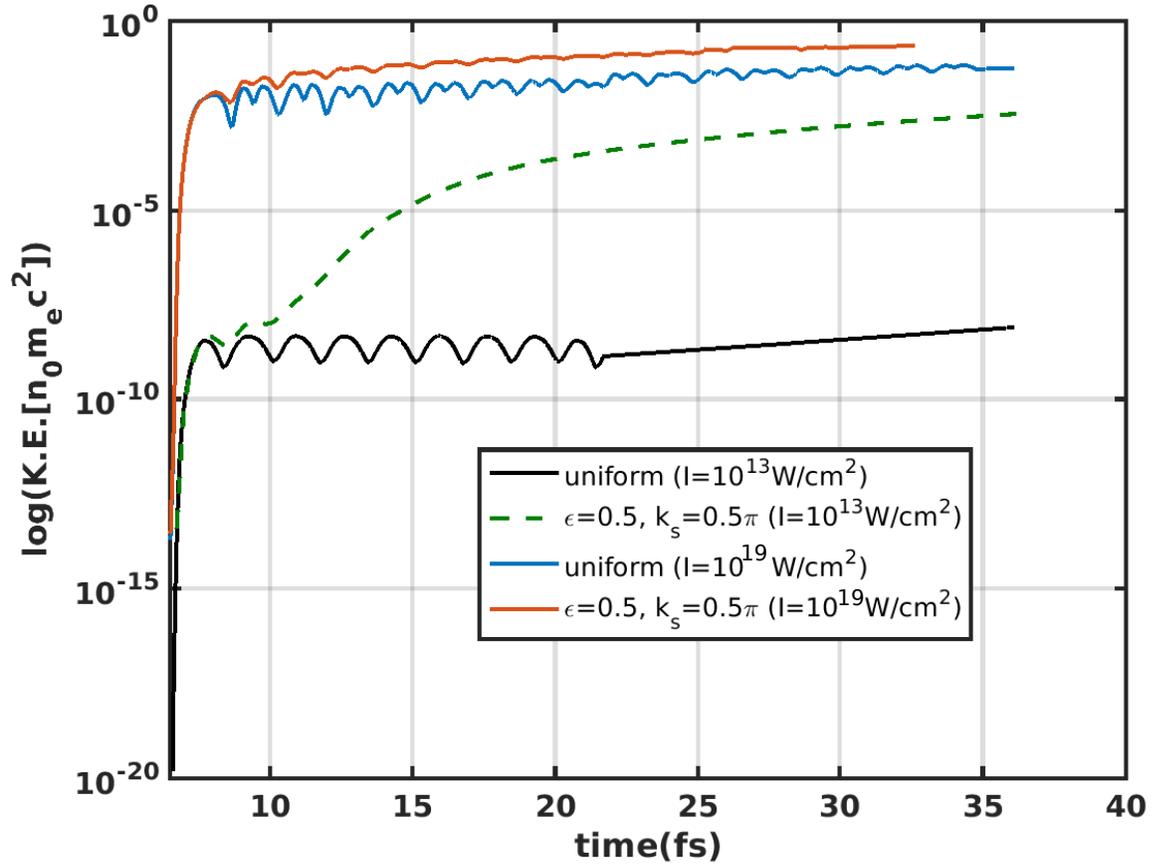}
               \caption{ Total kinetic energy of electrons [in unit of $n_0m_ec^2$] for various parameters of
ripple for intensity I$=1\times10^{13}W/cm^{2}$ and I$=1\times10^{19}W/cm^{2}$}  
                \label{fig4}
        \end{figure}

\begin{figure}[!htb]
               \includegraphics[width=\textwidth]{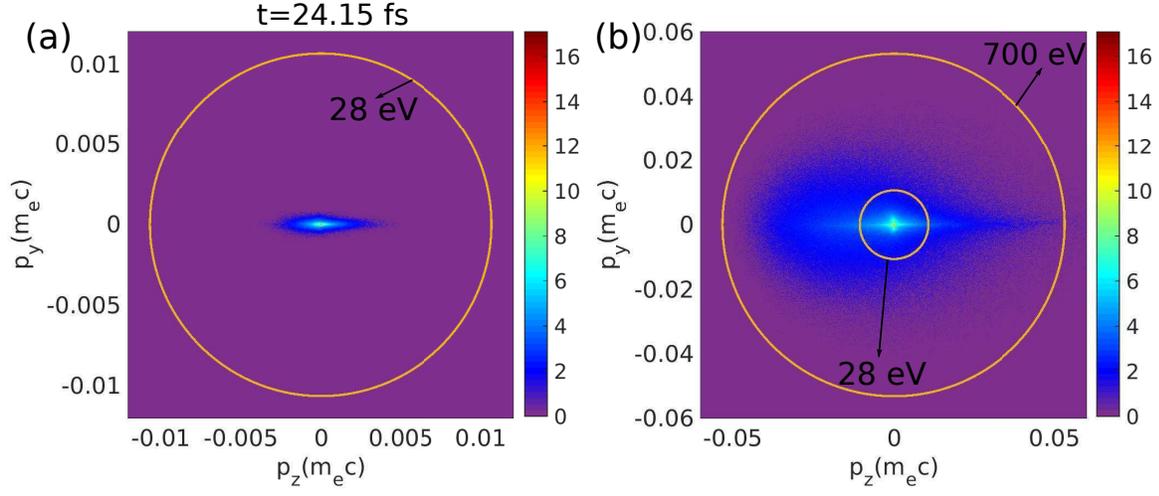}
              \caption{ The $p_zp_y$ [in unit of $m_ec$] phase space of electrons for laser intensity I$=1\times10^{13}W/cm^{2}$ at time t$=24.15$ fs, 
             the circles (yellow color) represent the energy curve with 28 eV and 700 eV respectively
             : (a) In uniform target case, momenta is confined within circle (b) In structured target ($\varepsilon=0.5$, $k_{s}=0.5\pi$), 
             the momenta spread as well as energy is higher which confirms that the structured target is comparatively better for laser absorption}  
                \label{fig5}
        \end{figure}

We now consider the absorption studies for the case when the laser intensity is weak $10^{13} W/cm^2$. 
For the planar homogeneous target, we expect that 
 the vacuum heating would be inoperative as the electric field vector of the laser (which is incident normal to the target)  is parallel to the surface. 
The intensity  of $10^{13} W/cm^2$ is weak and electrons remain non-relativistic. Therefore, the 
 $\vec{J} \times \vec{B}$ mechanism is insignificant for this case. This is indeed observed as can be seen from Fig.~\ref{fig4}, where we have  plotted total kinetic energy of the electrons as a function of time. For the same laser intensity and all other conditions 
 when the target is taken to be inhomogeneous the total kinetic energy of electrons registers a significant growth with time as witnessed from the 
 dashed green line of the same Fig.~\ref{fig4}. 
 We also  denoted the electron number distribution (on a color log scale) as the function of $p_y$ and $p_z$ at $t = 24.15$ fs in Fig.~\ref{fig5} 
  for the both planar homogeneous and structured target cases and we also draw the energy circles (with yellow color) on $p_y-p_z$ plane. For planar homogeneous target, in Fig.~\ref{fig5}(a) we see that the
  spread of electron number distribution is very low and even do not cross the energy circle of 28 eV. 
  However, for structured inhomogeneous target
  the momenta spread of electron is around 700 eV (see Fig.~\ref{fig5}(b)) which is comparatively much higher than homogeneous. Thus it is clearly evident that the laser radiation absorption is considerably better for the structured target. 
 
\begin{figure}[!htb]
               \includegraphics[width=\textwidth]{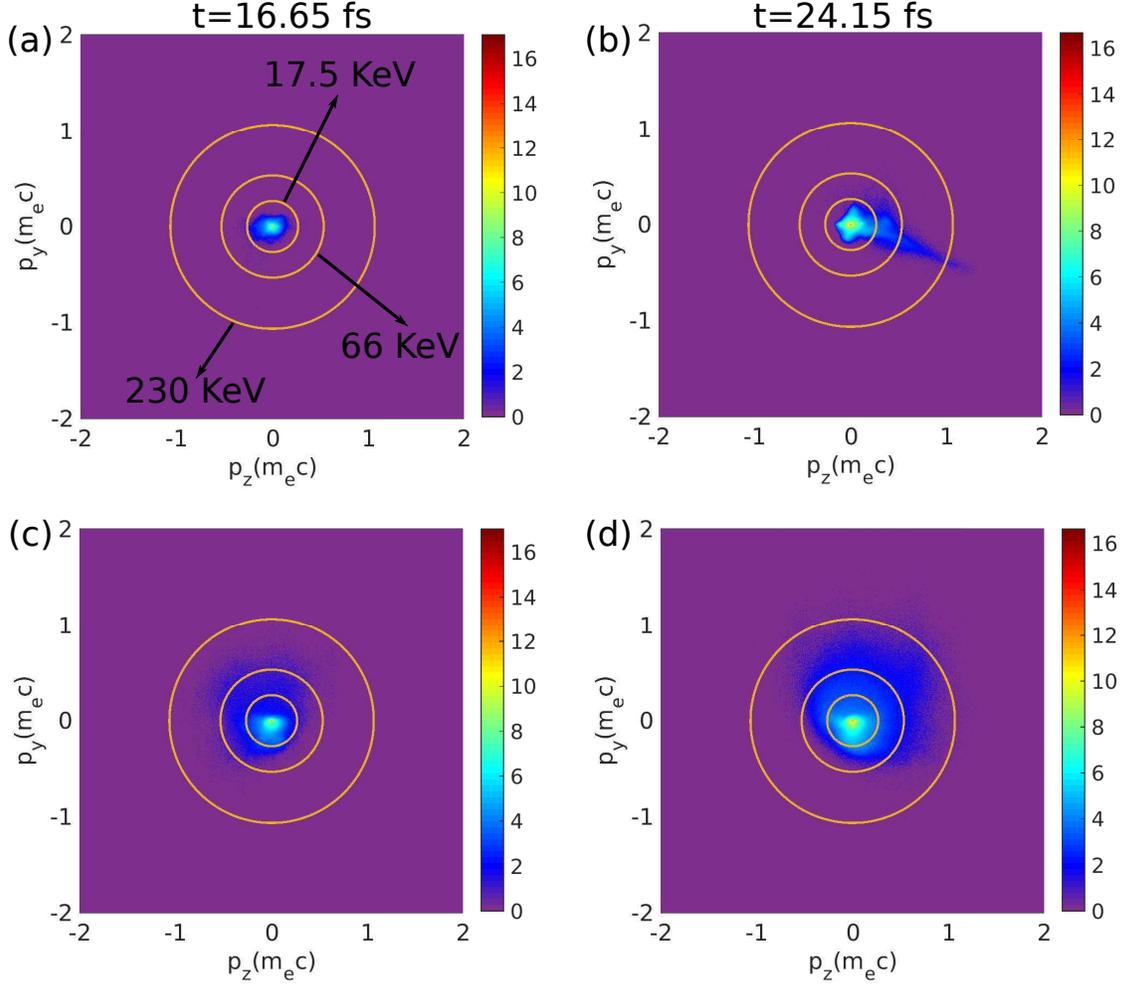}
              \caption{The $p_zp_y$ [in unit of $m_ec$] phase space of electrons for laser intensity I$=1\times10^{19}W/cm^{2}$, 
             the circles (yellow color) represent the energy curve with 17.5, 66 and 230 KeV respectively
             : (a) \& (b) The momentum at time t$=16.65$ and t$=24.15$ fs respectively for uniform target case where collimation in momenta can be seen (c) \& (d) In structured target ($\varepsilon=0.5$, $k_{s}=0.5\pi$), 
             the momenta spread as well as energy is higher which confirms that the structured target is comparatively better for laser absorption}  
                \label{fig6}
        \end{figure}

 We have repeated the study at higher relativistic intensity of the laser (viz. I$=10^{19} W/cm^2$). In this case, even for the homogeneous target there is a significant increase in the electron energy. 
 However, for the structured case the acquired energy is still considerably higher. This is evident  from Fig.~\ref{fig4} which shows the growth of the total electron kinetic energy as 
 well as from the plots of Fig.~\ref{fig6}(c) and (d) which show the momenta spread in the $p_y$ vs. $p_z$ plane. In Fig.~\ref{fig6}, the time evolution of the electron number distribution as function of momenta for both  homogeneous as well as inhomogeneous structured target is shown. For homogeneous planar
 target, we see from Fig.~\ref{fig6}(a) and(b) that electron number distribution at time $t=16.65$ fs have maximum energy around 17.5 KeV and at later time $t=24.15$ fs,  few electrons have energy around 230 KeV with narrow spread.
 However, in the case of inhomogeneous structured target, at time  $t=16.65$ fs the electrons have more energy around 66 KeV as well as more spread (see Fig.~\ref{fig6}(c))compare to planar homogeneous target. Furthermore, at later time $t=24.15$ fs, the most of the electrons have energy around 230 KeV (see Fig.~\ref{fig6}(d)) which indicates that laser absorption is more in structured target.
           
\begin{figure}[!htb]
               \includegraphics[width=\textwidth]{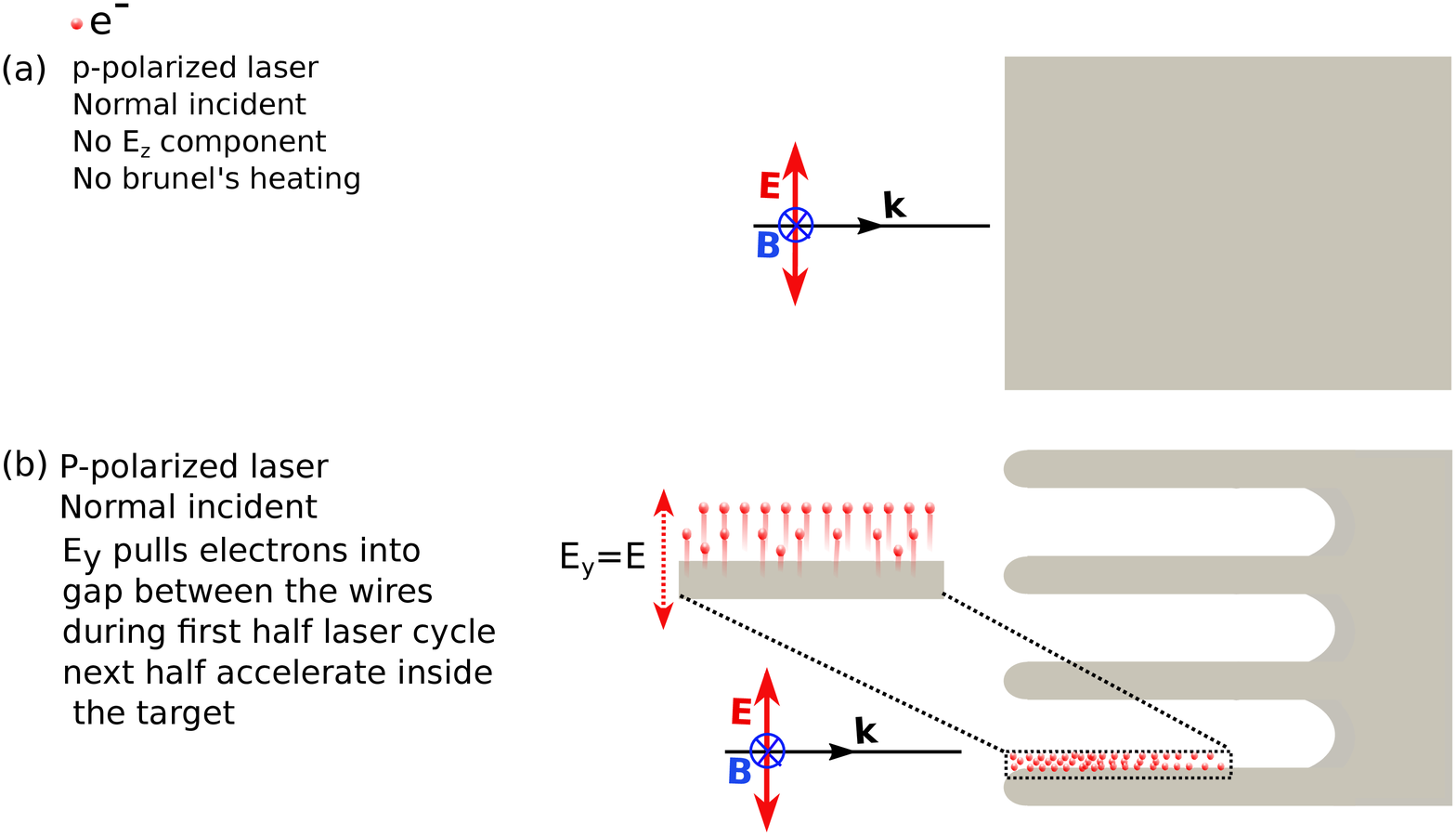}
               \caption{ Sketch of vacuum heating mechanism (a) In normal incidence for homogeneous planar target, there is no electric field normal to plasma surface. Therefore, there is no vacuum heating (Brunel's mechanism). 
             (b) However, for structured nano-wire target, in normal incidence case, there is electric field normal to side plasma surface which satisfies the vacuum heating condition and increases the absorption of laser radiation compare to conventional vacuum heating (Brunel's mechanism).}
        \label{fig7}
       \end{figure}

The understanding of the enhanced absorption can be understood from the schematic cartoon plots of Fig.~\ref{fig7}. 
When a laser is normally incident on a homogeneous target there is no component of electric field which can drag the electrons out 
in the vacuum as shown in Fig.~\ref{fig7}(a). However, when the surface is corrugated as shown in Fig.~\ref{fig7}(b) even at normal incidence the electric field component $E_y$ can drag the electrons 
out in the  vacuum like low density region  from  the high  plasma density region. Furthermore, the laser fields in this case also access an increased surface area. 
It is now of interest to know how does the absorption depend on the inhomogeneity scale length. 
 and the inhomogeneity  amplitude (which defines the density disparity). The next section provides the details of this study. 
\section{Dependence of absorption on structure parameters}
\begin{figure}[!htb]
               \includegraphics[width=\textwidth]{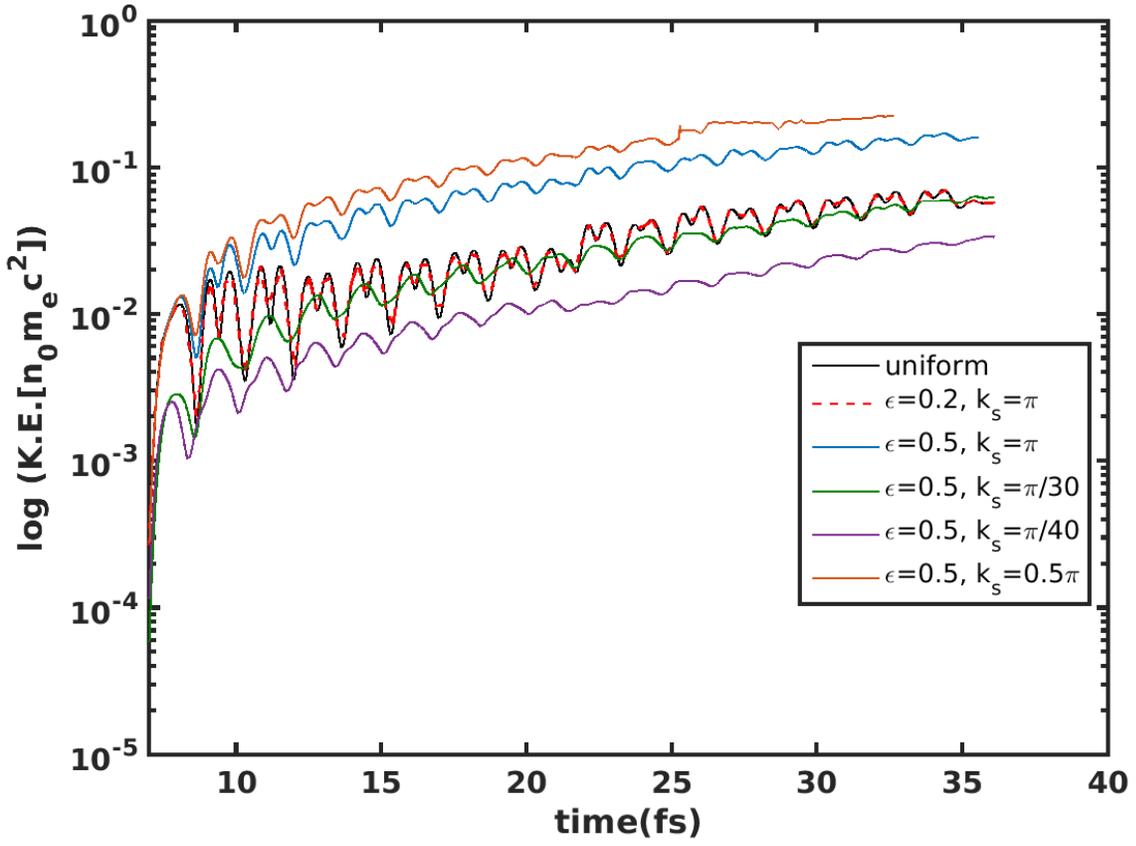}
               \caption{ Total kinetic energy of electrons [in unit of $n_0m_ec^2$] for various parameters of
ripple for intensity I$=1\times10^{19}W/cm^{2}$}  
                \label{fig8}
        \end{figure}

  We have considered various inhomogeneity scale lengths and amplitude of the plasma density profile and studied  the total kinetic energy acquired by the electrons 
 as a function of time. This has 
 been shown in Fig.~\ref{fig8}. This figure clearly shows that the total kinetic energy of electrons is largest  only for an  intermediate value of 
 ($k_s=0.5\pi$ ($l_s=4c/\omega_p$) of the  inhomogeneity scale length.  
 Both increasing and/or decreasing the inhomogeneity  scale length 
results in reduced absorption. This can be understood by realizing that for the 
 laser intensity I$=1\times 10^{19}W/cm^{2}$ considered for this study, the  distance by which the electrons can be dragged out  along the y-direction $y_{osc}=eE_L/\gamma 
 m_e\omega_L^2$ in one laser cycle is also 
 about $4c/\omega_p$ where $E_L$ and $\omega_L$ are laser electric field and laser frequency respectively. 
Thus, when the inhomogeneity scale length is much sharper than $y_{osc}$ the dragged electron would enter another high density 
region on the other side in one laser cycle as the spacing between intervening high and low densities are very small. The electrons thus would not experience the 
vacuum like region important for the Brunel mechanism to be operative.  When the inhomogeneity scale length is much broader than the 
$y_{osc}$, even then the electrons will not experience the vacuum region. Thus the optimum scale length is when  $y_{osc}$ is comparable with the inhomogeneity scale length. 
 
  It should also be noted from Fig.~\ref{fig8} that there is an amplitude dependence also in the absorption rate. A high amplitude of inhomogeneity seems to do better 
for absorption as the contrast between low and high density region is better for this case. 
\section{Summary and Discussion }
We have carried out the 2D PIC simulations of the short pulse intense laser interaction with the rippled pre-ionized plasma target. 
It is shown that the laser absorption is better for structured targets for both relativistic as well as non-relativistic laser intensities.  
It is shown that this happens as a result of providing increased surface area for the Brunel's mechanism of vacuum heating. Furthermore, 
even at normal incidence, there is no component of electric field normal to the surface to drag the electrons out in the vacuum region for a normal 
homogeneous target.  The structured target, on the other hand, provides a geometry for the electric field to be normal to the imposed inhomogeneity. 
Thus, the electrons can be dragged out 
 from high to low density regions for a vacuum heating like mechanism to be operative. 

There have been experiments \cite{g.r.k} which have reported increased propagation distance  of energetic  electron beams  generated by lasers in  nano structured targets. 
The explanation for the phenomena has been provided on the basis of suppression of Weibel instability  by the structured target which aids the unhindered propagation 
of the electrons in the medium. It appears  that in such experiments the enhanced generation of energetic electrons can also  play an additional  role in aiding the 
propagation over long distance. 

It is interesting to note that the structured targets play important roles both in enhanced generation of energetic electrons as well as 
aiding its propagation through plasma medium over long distances by suppressing Weibel like instabilities which would otherwise be greatly detrimental. 
 
{\bf{Acknowledgement:} }The authors are grateful to the late Prof. Predhiman Kaw and Dr. Kartik Patel for useful discussions.
\clearpage
\newpage
 \bibliographystyle{unsrt}

\newpage

\end{document}